\documentclass[12pt]{article}
\newcommand{\sect}[1]{\setcounter{equation}{0}\section{#1}}

\renewcommand{\appendix}{\setcounter{section}{0}
\renewcommand{\thesection}{\Alph{section}}}

\def\a{\alpha}
\def\b{\beta}
\def\g{\gamma}

\def\e{\epsilon}

\def\t{\theta}

\def\s{\sigma}

\def\be{\begin{equation}}
\def\ee{\end{equation}}
\def\ba{\begin{eqnarray}}
\def\ea{\end{eqnarray}}

\newcommand{\nn}{\nonumber\\}
\newcommand{\no}{\nonumber}
\begin{document}
\renewcommand{\thefootnote}{\fnsymbol{footnote}}

\newpage
\setcounter{page}{0}
\pagestyle{empty}

\begin{center}
{\Large{\bf Tensor extension of the Poincar\'e algebra\\}}
\vspace{1cm}
{\bf Dmitrij V. Soroka\footnote{E-mail: dsoroka@kipt.kharkov.ua} and
Vyacheslav A. Soroka\footnote{E-mail: vsoroka@kipt.kharkov.ua}}
\vspace{1cm}\\
{\it Kharkov Institute of Physics and Technology,
61108 Kharkov, Ukraine}\\
\vspace{1.5cm}
\end{center}
\begin{abstract}
A tensor extension of the Poincar\'e algebra is proposed for the arbitrary 
dimensions. Casimir operators of the extension are constructed. A possible 
supersymmetric generalization of this extension is also found in the 
dimensions $D=2,3,4$.



\end{abstract}

\newpage
\pagestyle{plain}
\renewcommand{\thefootnote}{\arabic{footnote}}
\setcounter{footnote}0

\sect{Introduction}

There are many examples for the tensor (semi-central) extensions of the 
super-Poincar\'e algebra (see, for 
example,~\cite{dafr,holpr,z,hugpol,vst,agit,ds}). However, there also exists 
the tensor extension of the Poincar\'e algebra itself. In the present paper we 
give the example of such an extension with the help of the second rank tensor 
generator. A consideration of such an extension may have a sense, since it is 
homomorphic to the Poincar\'e algebra. Moreover, the contraction of the 
extended algebra leads also to the Poincar\'e algebra. It is interesting 
enough that the momentum square Casimir operator for the Poincar\'e algebra 
under this extension ceases to be the Casimir operator and it is generalized 
by adding the term containing linearly the angular momentum\footnote{Note that 
this reminds the relation for the Regge trajectory, which connects the mass 
square with the angular momentum.}. Due to this fact, an irreducible 
representation of the extended algebra\footnote{Concerning the irreducible 
unitary representations of the extended Poincar\'e group in $(1+1)$ dimensions 
see, for example,~\cite{mr}.} has to contain the fields of the different 
masses. This extension with non-commuting momenta has also something in common 
with the ideas of the papers~\cite{sn,ya,hl} and with the non-commutative 
geometry idea~\cite{c}.
It is also shown that for the dimensions $D=2,3,4$ the extended 
Poincar\'e algebra allows a supersymmetric generalization.

\sect{Extension of the Poincar\'e algebra}

The Poincar\'e algebra for the components of the rotations $M_{ab}$ and 
translations $P_a$ in $D$ dimensions
\ba
[M_{ab},M_{cd}]=(g_{ad}M_{bc}+g_{bc}M_{ad})-(c\leftrightarrow d),\no
\ea
\ba
[M_{ab},P_c]=g_{bc}P_a-g_{ac}P_b,\no
\ea
\ba\label{2.1}
[P_a,P_b]=0
\ea
can be extended with the help of the tensor semi-central generator $Z_{ab}$
in the following way:
\ba
[M_{ab},M_{cd}]=(g_{ad}M_{bc}+g_{bc}M_{ad})-(c\leftrightarrow d),\no
\ea
\ba
[M_{ab},P_c]=g_{bc}P_a-g_{ac}P_b,\no
\ea
\ba
[P_a,P_b]=Z_{ab},\no
\ea
\ba
[M_{ab},Z_{cd}]=(g_{ad}Z_{bc}+g_{bc}Z_{ad})-(c\leftrightarrow d),\no
\ea
\ba
[P_a,Z_{bc}]=0,\no
\ea
\ba\label{2.2}
[Z_{ab},Z_{cd}]=0.
\ea
By taking a set of the generators $Z_{ab}$ as a homomorphism kernel, we obtain
that the extended Poincar\'e algebra (\ref{2.2}) is homomorphic to the usual
Poincar\'e algebra (\ref{2.1}). Under a contraction ${Z_{ab}\to 0}$ the algebra
(\ref{2.2}) also goes to the Poincar\'e algebra (\ref{2.1}).

Casimir operators of the extended Poincar\'e algebra are
\ba\label{2.3}
Z_{a_1a_2}Z^{a_2a_3}\cdots Z_{a_{2k-1}a_{2k}}Z^{a_{2k}a_1},\quad(k=1,2,\ldots);
\ea
\ba\label{2.4}
P^{a_1}Z_{a_1a_2}Z^{a_2a_3}&\cdots& 
Z_{a_{2k-1}a_{2k}}Z^{a_{2k}a_{2k+1}}P_{a_{2k+1}}\cr\nn
+Z^{aa_1}Z_{a_1a_2}&\cdots&
Z_{a_{2k-1}a_{2k}}Z^{a_{2k}a_{2k+1}}M_{a_{2k+1}a},\quad(k=0,1,2,\ldots);
\ea
\ba\label{2.5}
\e^{abcd}Z_{ab}Z_{cd},
\ea
where $\e^{abcd}$, $\e^{0123}=1$ is the totally antisymmetric Levi-Civita 
tensor. In particular, there is a Casimir operator generalizing the momentum 
square
\ba\label{2.6}
P^aP_a+Z^{ab}M_{ba},
\ea
which indicates that an irreducible representation of the extended algebra
contains the fields having the different masses. Note that for the extended 
algebra there is no generalization of the Pauli-Lubanski vector of the 
Poincar\'e algebra. The expressions (\ref{2.3}) and (\ref{2.4}) for the Casimir
operators are valid for the extended Poincar\'e algebra (\ref{2.2}) in the 
arbitrary dimensions $D$, but the expression (\ref{2.5}) is only true for 
$D=4$.

Note that in the case of the extended two-dimensional Poincar\'e algebra the 
Casimir operators (\ref{2.3}) and (\ref{2.4}) can be expressed
\ba
Z_{a_1a_2}Z^{a_2a_3}\cdots Z_{a_{2k-1}a_{2k}}Z^{a_{2k}a_1}=2Z^{2k},\no
\ea
\ba
P^{a_1}Z_{a_1a_2}Z^{a_2a_3}&\cdots& 
Z_{a_{2k-1}a_{2k}}Z^{a_{2k}a_{2k+1}}P_{a_{2k+1}}\cr\nn
+Z^{aa_1}Z_{a_1a_2}&\cdots&
Z_{a_{2k-1}a_{2k}}Z^{a_{2k}a_{2k+1}}M_{a_{2k+1}a}=Z^{2k}(P^aP_a+Z^{ab}M_{ba})
\no
\ea
as degrees of the following generating Casimir operators:
\ba
Z={1\over2}\e^{ab}Z_{ab},\no
\ea
\ba
P^aP_a+Z^{ab}M_{ba},\no
\ea
where $\e^{ab}=-\e^{ba}$, $\e^{01}=1$ is the completely antisymmetric 
two-dimensional Levi-Civita tensor. In the case of the extended 
three-dimensional Poincar\'e algebra these Casimir operators can be expressed
\ba
Z_{a_1a_2}Z^{a_2a_3}\cdots Z_{a_{2k-1}a_{2k}}Z^{a_{2k}a_1}=2(Z^aZ_a)^k,\no
\ea
\ba
P^{a_1}Z_{a_1a_2}Z^{a_2a_3}&\cdots& 
Z_{a_{2k-1}a_{2k}}Z^{a_{2k}a_{2k+1}}P_{a_{2k+1}}\cr\nn
+Z^{aa_1}Z_{a_1a_2}&\cdots&
Z_{a_{2k-1}a_{2k}}Z^{a_{2k}a_{2k+1}}M_{a_{2k+1}a}\cr\nn
&=&(Z^aZ_a)^k(P^aP_a+Z^{ab}M_{ba})-(Z^aZ_a)^{k-1}(P^aZ_a)^2
\no
\ea
in terms of the following generating Casimir operators:
\ba
Z^aZ_a,\no
\ea
\ba
P^aP_a+Z^{ab}M_{ba},\no
\ea
\ba
P^aZ_a,\no
\ea
where
\ba
Z^a={1\over2}\e^{abc}Z_{bc}\no
\ea
and $\e^{abc}$, $\e^{012}=1$ is the totally antisymmetric three-dimensional
Levi-Civita tensor.
In the case of the extended $D$-dimensional ($D\ge4$) Poincar\'e algebra  
the Casimir operators (\ref{2.3}) and (\ref{2.4}) can not be expressed in 
terms of the finite number of the generating Casimir operators.

Generators of the left shifts, acting on the function $f(y)$ with a group 
element $G$,
\ba
[T(G)f](y)=f(G^{-1}y),\quad y=(x^a,z^{ab})\no
\ea
have the form
\ba
P_a=-\left({\partial\over\partial x^a}+
{1\over2}x^b{\partial\over\partial z^{ab}}\right),\no
\ea
\ba
Z_{ab}=-{\partial\over\partial z^{ab}},\no
\ea
\ba
M_{ab}=x_a{\partial\over\partial x^b}-x_b{\partial\over\partial x^a}
+{z_a}^c{\partial\over\partial z^{bc}}-{z_b}^c{\partial\over\partial z^{ac}}
+S_{ab},\no
\ea
where coordinates $x^a$ correspond to the translation generators $P_a$,
coordinates $z^{ab}$ correspond to the generators $Z_{ab}$ and 
$S_{ab}$ is a spin operator.

On the other hand, generators of the right shifts
\ba
[T(G)f](y)=f(yG)\no
\ea
have the form
\ba
D\mathrel{\mathop=^{\rm def}}
{P_a}^r={\partial\over\partial x^a}-
{1\over2}x^b{\partial\over\partial z^{ab}},\no
\ea
\ba
{Z_{ab}}^r=-Z_{ab}={\partial\over\partial z^{ab}}.\no
\ea
Note that the algebra
\ba
[M_{ab},M_{cd}]=(g_{ad}M_{bc}+g_{bc}M_{ad})-(c\leftrightarrow d),\no
\ea
\ba
[M_{ab},P_c]=g_{bc}P_a-g_{ac}P_b,\no
\ea
\ba
[M_{ab},D_c]=g_{bc}D_a-g_{ac}D_b,\no
\ea
\ba
[P_a,P_b]=Z_{ab},\no
\ea
\ba
[D_a,D_b]=-Z_{ab},\no
\ea
\ba
[P_a,D_b]=0,\no
\ea
\ba
[M_{ab},Z_{cd}]=(g_{ad}Z_{bc}+g_{bc}Z_{ad})-(c\leftrightarrow d),\no
\ea
\ba
[P_a,Z_{bc}]=0,\no
\ea
\ba
[D_a,Z_{bc}]=0,\no
\ea
\ba
[Z_{ab},Z_{cd}]=0,\no
\ea
formed by the generators $M_{ab}$, $P_a$, $D_a$ and $Z_{ab}$, has as  
Casimir operators the operators (\ref{2.3}) and the following operators:
\ba
(P-D)^{a_1}Z_{a_1a_2}Z^{a_2a_3}&\cdots& 
Z_{a_{2k-1}a_{2k}}Z^{a_{2k}a_{2k+1}}(P+D)_{a_{2k+1}}\cr\nn
+Z^{aa_1}Z_{a_1a_2}&\cdots&
Z_{a_{2k-1}a_{2k}}Z^{a_{2k}a_{2k+1}}M_{a_{2k+1}a},\quad(k=0,1,2,\ldots).\no
\ea

\sect{Supersymmetric generalization}

The extended Poincar\'e algebra (\ref{2.2}) admits the following supersymmetric
generalization:
\ba
\{Q_\a,Q_\b\}=-d(\s^{ab}C)_{\a\b}Z_{ab},\no
\ea
\ba
[M_{ab},Q_\a]=-(\s_{ab}Q)_\a,\no
\ea
\ba
[P_a,Q_\a]=0,\no
\ea
\ba
[Z_{ab},Q_\a]=0\no
\ea
with the help of the super-translation generators
\ba
Q_\a=-\left[{\partial\over\partial {\bar\t}^{\a}}+
{d\over2}(\s^{ab}\t)_\a{\partial\over\partial z^{ab}}\right],\no
\ea
where $\t=C\bar\t$ is a Majorana Grassmann spinor, $C$ is a charge conjugation 
matrix, $d$ is some constant and $\s_{ab}={1\over4}[\g_a,\g_b]$.

The rotation generators acquire the terms depending on the Grassmann variables 
$\t_\a$
\ba
M_{ab}=x_a{\partial\over\partial x^b}-x_b{\partial\over\partial x^a}
+{z_a}^c{\partial\over\partial z^{bc}}-{z_b}^c{\partial\over\partial z^{ac}}
-(\s_{ab}\t)_\a{\partial\over\partial\t_\a}+S_{ab},\no
\ea
whereas the expressions for the translations $P_a$ and tensor generator 
$Z_{ab}$ remain unchanged.

The validity of the Jacobi identities
\ba
[P_a,\{Q_\a,Q_\b\}]=\{Q_\a,[P_a,Q_\b]\}+\{Q_\b,[P_a,Q_\a]\}\no
\ea
and
\ba
[M_{ab},\{Q_\a,Q_\b\}]=\{Q_\a,[M_{ab},Q_\b]\}+\{Q_\b,[M_{ab},Q_\a]\}\no
\ea
for the supersymmetric generalization of the extended Poincar\'e algebra
(\ref{2.2}) verified for the dimensions $D=2,3,4$ with the use of the symmetry
properties of the matrices $C$ and $\g_aC$ and the relations 
(\ref{A.1})--(\ref{A.3}) of the Appendix.

One of the generating Casimir operator in the dimensions $D=2,3$ is generalized
 into the following form:
\ba\label{3.1}
P^aP_a+Z^{ab}M_{ba}-{1\over2d}Q_\a(C^{-1})^{\a\b}Q_\b,
\ea
while the form of the rest generating Casimir operators in these dimensions
are not changed.
Note that in the case $D=3$ there is also the following Casimir operator:
\ba
Z^aQ_\a(C^{-1}\g_a)^{\a\b}Q_\b.\no
\ea

One of the simplest Casimir operator (\ref{2.6}) in $D=4$
is also generalized into the form (\ref{3.1}). The supersymmetric 
generalization of the 
more complicated Casimir operators in the four-dimensional case has the 
following structure:
\ba
P^aZ_{ab}Z^{bc}P_c+Z^{ab}Z_{bc}Z^{cd}M_{da}
+{2\over5d}Q_\a(C^{-1}\s^{ab}Z_{ab}\s^{cd}Z_{cd})^{\a\b}Q_\b
+{1\over2d}Z^{ab}Z_{ab}Q_\a(C^{-1})^{\a\b}Q_\b,\no
\ea
\ba
P^aZ_{ab}Z^{bc}Z_{cd}Z^{de}P_e&+&Z^{ab}Z_{bc}Z^{cd}Z_{de}Z^{ef}M_{fa}\cr\nn
&+&{2\over5d}Q_\a\left[C^{-1}\s^{ab}Z_{ab}\s^{cd}\left(Z_{ce}Z^{ef}Z_{fd}
+{3\over10}Z^{gh}Z_{hg}Z_{cd}\right)\right]^{\a\b}Q_\b\cr\nn
&-&{1\over20d}\left[7Z_{ab}Z^{bc}Z_{cd}Z^{da}
+3(Z^{ef}Z_{fe})^2\right]Q_\a(C^{-1})^{\a\b}Q_\b.\no
\ea
An algorithm for the construction of the supersymmetric generalization of the 
Casimir operators (\ref{2.4}) is obvious and based on the use of the following 
commutation relations:
\ba
\left[{1\over2d}Q_\a(C^{-1})^{\a\b}Q_\b,Q_\g\right]=Z^{ab}(\s_{ab}Q)_\g,\no
\ea
\ba
\left[{2\over5d}Q_\a(C^{-1}\s^{ab}Z_{ab}\s^{cd}\tilde Z_{cd})^{\a\b}Q_\b,
Q_\g\right]=\biggl(Z^{ab}Z_{bc}\tilde Z^{cd}
&+&{7\over10}Z_{bc}\tilde Z^{cb}Z^{ad}\cr\nn
&+&{3\over10}Z_{bc}Z^{cb}\tilde Z^{ad}\biggr)(\s_{ad}Q)_\g,\no
\ea
where
\ba
\tilde Z^{ab}=Z^{aa_1}Z_{a_1a_2}\cdots Z_{a_{2k-1}a_{2k}}Z^{a_{2k}b},\quad
(k=0,1,\ldots).\no
\ea

\sect{Conclusion}

Thus, in the present paper we proposed the extension of the Poincar\'e 
algebra with the help of the second rank tensor generator. Casimir operators 
for the extended algebra are constructed. The form of the Casimir operators 
indicate that an irreducible representation of the extended algebra contains
the fields with the different masses. 
A consideration is performed for the arbitrary dimensions $D$.
A possible supersymmetric generalization 
of the extended Poincar\'e algebra is also given for the  
particular cases with the dimensions $D=2,3,4$.

It would be interesting to find the spectra of the Casimir operators and to 
construct the models based on the extended Poincar\'e algebra. The work in 
this direction is in progress.

\section*{Acknowledgments}

One of the authors (V.A.S.) would like to thank B.A. Dubrovin for useful 
discussions. V.A.S. is sincerely grateful to L. Bonora for fruitful discussions
and for kind hospitality at SISSA/ISAS (Trieste), where the main part of 
this work has been performed.

\appendix
\sect{Appendix}

As a real (Majorana) representation for the two-dimensional $\g$-matrices and 
charge conjugation matrix $C$ we adopt
\ba
\g^0=C=-C^T=-i\s_2,\quad\g^1=\s_1,\quad\g_5={1\over2}\e^{ab}\g_a\g_b=\s_3;\no
\ea
\ba
\{\g_a,\g_b\}=2g_{ab},\quad g_{11}=-g_{00}=1,\quad C^{-1}\g_aC=-{\g_a}^T,\no
\ea
where $\s_i$ are Pauli matrices. The matrices $\g_a$ satisfy the relations
\ba
\g_a\g_5=\e_{ab}\g^b,\quad\g_a\g_b=g_{ab}-\e_{ab}\g_5. \label{A.1}
\ea

For the Majorana three-dimensional $\g$-matrices and charge conjugation matrix 
$C$ we take
\ba
\g^0=C=-C^T=-i\s_2,\quad\g^1=\s_1,\quad\g^2=\s_3;\no
\ea
\ba
\{\g_a,\g_b\}=2g_{ab},\quad g_{11}=g_{22}=-g_{00}=1,\quad 
C^{-1}\g_aC=-{\g_a}^T,
\no
\ea
The matrices $\g_a$ obey the relations
\ba
\g_a\g_b=g_{ab}-\e_{abc}\g^c. \label{A.2}
\ea

At last, the real four-dimensional $\g$-matrices and matrix $C$ are
\ba
\g^0=C=-C^T=-i\left(\begin{array}{cc}0&\s_2\\
\s_2&0\end{array}\right),\quad
\g^1=\left(\begin{array}{cc}\s_3&0\\
0&\s_3\end{array}\right),\no
\ea
\ba
\g^2=i\left(\begin{array}{cc}0&\s_2\\
-\s_2&0\end{array}\right),\quad
\g^3=-\left(\begin{array}{cc}\s_1&0\\
0&\s_1\end{array}\right),
\no
\ea
\ba
\{\g_a,\g_b\}=2g_{ab},\quad g_{11}=g_{22}=g_{33}=-g_{00}=1,\quad 
C^{-1}\g_aC=-{\g_a}^T,\quad\g_5={1\over4}\e^{abcd}\g_a\g_b\g_c\g_d.\no
\ea
The matrices $\g_a$ and $\s_{ab}$ meet the relations
\ba
\g_a\s_{bc}={1\over2}\e_{abcd}\g^d\g_5+{1\over2}(\g_cg_{ab}-\g_bg_{ac}),\quad
\s_{ab}\g_c={1\over2}\e_{abcd}\g^d\g_5+{1\over2}(\g_ag_{bc}-\g_bg_{ac}),\no
\ea
\ba
\s_{ab}\s_{cd}={1\over4}(g_{ad}g_{bc}-g_{ac}g_{bd}-\e_{abcd}\g_5)
+{1\over2}(\s_{ad}g_{bc}+\s_{bc}g_{ad}-\s_{ac}g_{bd}-\s_{bd}g_{ac}).\label{A.3}
\ea

\end{document}